\documentclass[12pt,preprint]{aastex}
\shorttitle{}
\shortauthors{Nesvorn\'y et al.}

\begin{document}
\baselineskip 19.pt

\title{Very Slow Rotators from Tidally Synchronized Binaries}

\author{David Nesvorn\'y$^1$, David Vokrouhlick\'y$^2$, William F. Bottke$^1$, Harold F. Levison$^1$,\\
William M. Grundy$^3$}
\affil{(1) Department of Space Studies, Southwest Research Institute,\\
1050 Walnut St., Suite 300, Boulder, CO, 80302, USA}
\affil{(2) Institute of Astronomy, Charles University,\\ 
V Hole\v{s}ovi\v{c}k\'ach 2, CZ--18000 Prague 8, Czech Republic}
\affil{(3) Lowell Observatory, 1400 W. Mars Hill Rd., Flagstaff, AZ 86001, USA}

\begin{abstract} 
A recent examination of K2 lightcurves indicates that $\sim$15\% of Jupiter Trojans have very slow 
rotation (spin periods $P_{\rm s}>100$ h). Here we consider the possibility that these bodies formed 
as equal-size binaries in the massive outer disk at $\sim$20-30 au. Prior to their implantation as 
Jupiter Trojans, tight binaries tidally evolved toward a synchronous state with $P_{\rm s} \sim P_{\rm b}$, 
where $P_{\rm b}$ is the binary orbit period. They may have been subsequently dissociated by impacts and 
planetary encounters with at least one binary component retaining its slow rotation. Surviving
binaries on Trojan orbits would continue to evolve by tides and spin-changing impacts over 4.5 Gyr. 
To explain the observed fraction of slow rotators, we find that at least $\sim$15-20\% of outer disk 
bodies with diameters $15<D<50$ km would have to form as equal-size binaries with $12\lesssim a_{\rm b}/R 
\lesssim 30$, where $a_{\rm b}$ is the binary semimajor axis and $R=D/2$. The mechanism proposed here could 
also explain very slow rotators found in other small body populations. 
\end{abstract}


\section{Slow rotators among Jupiter Trojans}

\begin{figure}
\epsscale{0.8}
\plotone{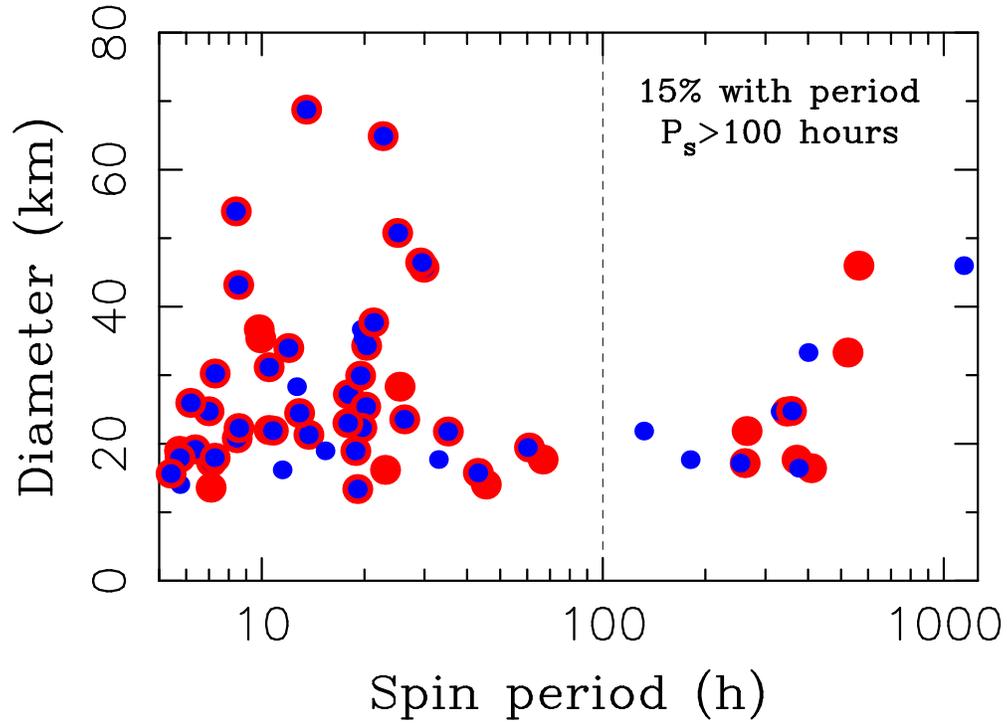}
\caption{The spin periods of Jupiter Trojans from Ryan et al. (2017) (bigger red dots) and Szab\'o et al. (2017)
(smaller blue dots). We excluded three cases with inconsistent period determinations (see the main text), 
leaving 53 objects. We included all cases where the period determination in one work was double of the one 
found in the other work. The period determinations in Ryan et al. and Szab\'o et al. precisely agree with each 
other in 40 cases (blue and red dots overlap in the plot).}
\label{period}
\end{figure}

Szab\'o et al. (2017) and Ryan et al. (2017) analyzed the {\it Kepler} telescope photometry during K2 
Campaign 6 from 2015 July 14 to September 30, and determined spin periods for 56 Jupiter Trojans. Thanks 
to a nearly continuous photometric coverage of all targets over a 78 day campaign this dataset 
is not biased against slow rotators. 
We cross-linked the two datasets to find that the results are generally consistent (Fig. \ref{period}). 
In two cases, (13185) Agasthenes and (65240) 2002 EU106, the estimated periods are significantly different 
(e.g., Ryan et al. finds $P_s=113$~h for Agasthenes, whereas Szab\'o et al. finds $P_s=11.6$ h, probably 
because this object has a very small lightcurve amplitude). We also discarded (39270) 2001 AH11, for which 
Szab\'o et al. gives a three times longer period than Ryan et al. In some cases, the periods differ by a 
factor of two due to ambiguity in the lightcurve phase folding. We include these cases in Fig. \ref{period} because 
they do not affect the classification of Trojans into fast and slow rotators. In total, 8 out of 53 Trojans 
($\sim$15\%) have very long spin periods ($P_{\rm s}>100$~h). 

Rotation of a small body is set by planetesimal formation processes (Johansen \& Lacerda 2010, Visser et al. 
2020), and evolves by impacts (Farinella et al. 1992) and radiation effects such as YORP (Vokrouhlick\'y et al. 
2003). Formation processes and impacts are expected to yield the Maxwellian distribution (e.g., 
Medeiros et al. 2018) with predominantly fast spins. The YORP effect was previously suggested to explain 
the excess of slow rotators among small asteroids (Pravec et al. 2008). We find it unlikely that the 
YORP effect can explain the {\it very} slow rotators in Fig. \ref{period}. Scaling from the detected YORP 
acceleration of asteroid Bennu (${\rm d}{\omega}/{\rm d}t=2.6 \times 10^{-6}$ deg d$^{-2}$; Nolan et al. 2019) to semimajor 
axis $a=5.2$ au and diameter $D=30$ km, we estimate that it would take $\sim$40 Gyr for YORP to change the 
rotational period from 15 to 500 hours (the YORP timescale scales with $a^2 D^2$; Vokrouhlick\'y et al. 2015). 
In addition, the two slowest rotators with $P_{\rm s}>500$ h are also the largest ((23958) 1998 VD30 with 
$D\simeq46$ km and (13366) 1998 US24 with $D\simeq33$ km), whereas YORP is the most effective for small bodies.  
Adding to that, (11351) Leucus, a $D \simeq 35$-km target of the {\it Lucy} mission (Levison et al. 2016), 
has $P_{\rm s}=513.7$ h (French et al. 2015).

\section{Implantation of Jupiter Trojans and binaries}

Current dynamical models suggest that Jupiter Trojans formed in the outer planetesimal disk at $\sim$20-30 au
and were implanted onto their present orbits after having a series of scattering 
encounters with the outer planets (Morbidelli et al. 2005, Nesvorn\'y et al. 2013; also see Pirani et al. 
2019). The orbital excitation during encounters can explain the high orbital inclinations of Trojans. 
The formation of Jupiter Trojans at $\sim$20-30 au is reinforced by their similarities (e.g., colors, size 
distribution) to the trans-Neptunian objects (TNOs; e.g., Fraser et al. 2014, Emery et al 2015, Wong \& Brown 
2016). 

TNOs can be classified into two dynamical categories: cold and hot. The cold population (also called the 
`cold classical' population), with semimajor axes $a=42$-47 au and $i<5^\circ$, is thought to have 
formed in situ at $>40$ au (e.g., Parker and Kavelaars 2010, Batygin et al. 2011). At least 30\% (Noll et al. 2008), 
and perhaps as much as 100\% (Fraser et al. 2017), of cold classicals formed as equal-size binaries (e.g., 
Goldreich et al. 2002, Nesvorn\'y et al. 2010). The equal-size binaries are relatively rare in the 
hot population (hot classicals, resonant and scattered; see Gladman et al. 2008 for definitions), probably 
because wide and thus observable binaries starting at $\sim$20-30 au were dissociated by impacts and planetary 
encounters prior to their implantation into the Kuiper belt (Petit \& Mousis 2004, Parker \& Kavelaars 2010). 

Some wide equal-size binaries survived in the hot population suggesting that the initial binary fraction 
at $\sim$20-30 au was high (Nesvorn\'y \& Vokrouhlick\'y 2019). The chances of survival are better for 
tight binaries that are more strongly bound together. These binaries are difficult to observationally 
detect in the Kuiper belt (Noll et al. 2008), but can be resolved closer in. For example, (617) Patroclus 
and Menoetius, a pair of 100-km class Jupiter Trojans, is a binary with $a_{\rm b} \simeq 670$ km 
(Marchis et al. 2006). The Patroclus-Menoetius binary evolved by tides into a synchronous state, where the 
orbital period matches spin periods of both components, $P_{\rm s} = P_{\rm b} = 103.5$~h (Mueller et al. 2010). 
It is not plotted in Fig. \ref{period}, but if it would, it would contribute to the group of very slow rotators 
with $P_{\rm s}>100$ h. We are therefore compelled to consider the possibility that the slow rotators among 
Jupiter Trojans are tidally synchronized binaries (surviving or dissociated).

\section{Tidal synchronization timescale}

For a binary with primary and secondary radii $R_1$ and $R_2$, densities $\rho_1=\rho_2$ ($=\rho$) and masses 
$m_1$ and $m_2$, the tidal evolution of secondary's spin is given by:
\begin{equation}
{1 \over \omega^*} {{\rm d} \omega \over {\rm d}t} = - \tau_\omega^{-1}\ ,
\end{equation}
where
\begin{equation}
\tau_\omega^{-1} = {15 \over 4} {k \over Q} \left({R_1 \over a}\right)^{9/2} n\ 
\label{omega}
\end{equation} 
is the synchronization timescale (Goldreich \& Sari 2009). Here, $\omega^*=(G m_2/R_2^3)^{1/2}$ is the breakup 
spin rate, $G$ is the gravitational constant, $k$ is the tidal Love number, $Q$ is the tidal quality factor, 
and $n$ is the orbital frequency. We assume that the spin is prograde and initially fast 
($\omega=\omega^* \gg n$). A similar equation holds for the evolution of primary's spin. The basic 
problem with using Eq. (\ref{omega}) for TNOs or Jupiter Trojans is that the $Q/k$ factor is unknown. 

\begin{figure}
\epsscale{0.8}
\plotone{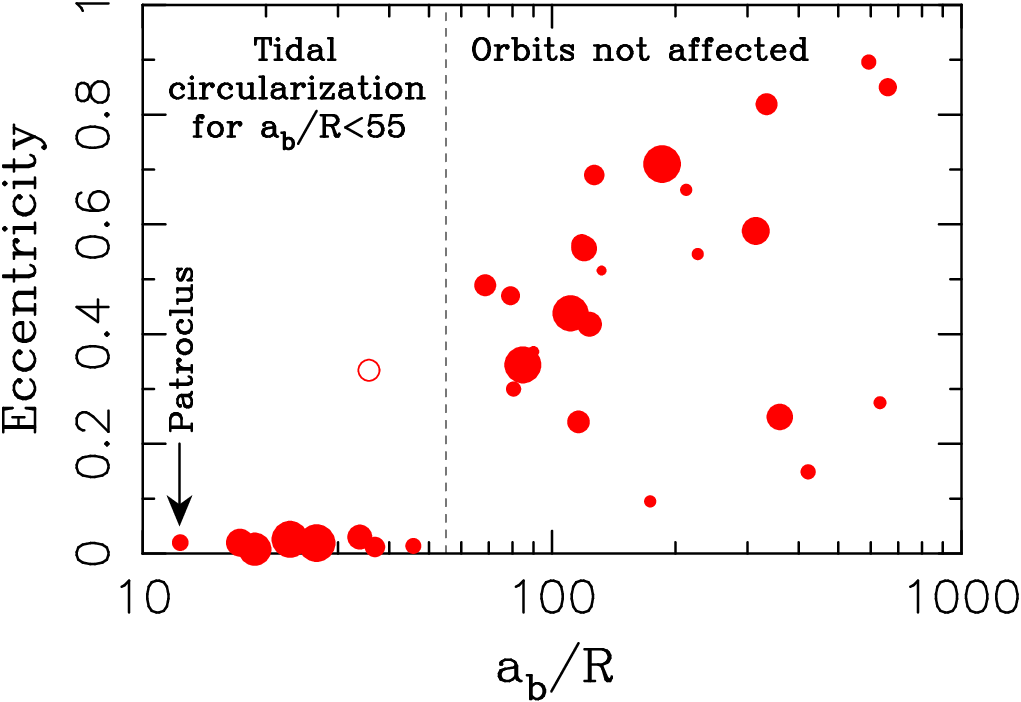}
\caption{Orbits of Kuiper belt binaries. We selected 36 binaries from Noll et al. (2019) with 
$R_1<150$ km (to remove dwarf planets) and $R_2/R_1 > 0.5$ (approximately equal size). The binary 
semimajor axis was normalized by the mean physical radius $R=(R_1+R_2)/2$, where the primary and secondary 
radii, $R_1$ and $R_2$, were taken from the Johnston's archive (Johnston 
2018; {\tt https://sbn.psi.edu/pds/resource/binmp.html}), Vilenius et al. (2014) and Stansberry 
et al. (2012). A similar plot can be obtained if the radii are computed from absolute magnitudes 
(Grundy et al. 2019, Noll et al. 2020) and albedo $p_{\rm V} \simeq 0.1$. The size of a symbol
is proportional to $R_1$ (the smallest symbol corresponds to $R_1=32$ km of 2000 CF105; the largest 
to $R_1=149$ km of 2002 XH91). The great majority of binaries shown here are cold classicals 
(30 in total). We do not show Centaur binary (42355) Typhon whose binary eccentricity $e_b=0.51$ was likely 
excited by planetary encounters (Nesvorn\'y \& Vokrouhlick\'y 2019). The circle symbol is a hot 
classical binary 2001 QC298 with a very low albedo $p_{\rm V}=0.034$ from Vilenius et al. (2014). 
It would shift to the right if the actual albedo is higher. The Patroclus-Menoetius binary with 
a circularized and double synchronous binary orbit is indicated by the arrow.}
\label{circ}
\end{figure}

We use observations of TNO binaries to infer this factor. The equal size, 100-km class TNO 
binaries show a clear trend of binary eccentricity $e_{\rm b}$ with separation (Fig.~\ref{circ}). 
The binaries with $a_{\rm b}/R<55$ have nearly circular orbits with $e_{\rm b}<0.03$, whereas the more
separated binaries have a wide range of eccentricities. We interpret this trend to be a consequence
of tidal circularization. The binary eccentricity evolution due to tides (Goldreich \& Sari 2009) is given 
by
\begin{equation}
{1 \over e_{\rm b}} {{\rm d} e_{\rm b} \over {\rm d}t} = - \tau_e^{-1}
\end{equation}
with the eccentricity damping timescale
\begin{equation}
\tau_e^{-1} = {k \over Q} n \left[
{21 \over 2} {m_1 \over m_2} \left({R_2 \over a}\right)^5
- {57 \over 8} {m_2 \over m_1} \left({R_1 \over a}\right)^5
\right] \ ,
\label{ecc}
\end{equation}
where we assume that the $k/Q$ factor is the same for primary and secondary. Eq. (\ref{ecc}) strictly 
applies only for $e \ll 1$. Here we use it as a guide for the timescale on which a binary orbit is
circularized.\footnote{We ignore the effect of tides on binary semimajor axis. This effect is small 
because there is not enough angular momentum contained, relative to the orbital momentum, in the initial 
spin (even if it is critical).}

\begin{figure}
\epsscale{0.8}
\plotone{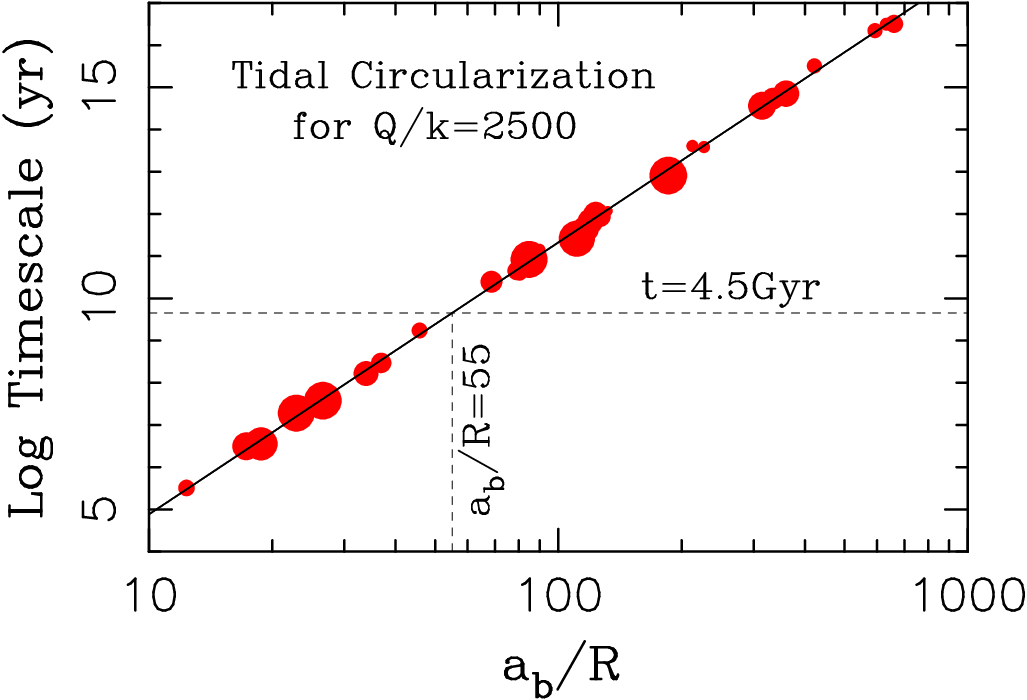}
\caption{The tidal circularization timescale $\tau_e$ of TNO binaries. We compute $\tau_e$ from Eq.~(\ref{ecc}) 
for TNO binaries shown in Fig. \ref{circ}, find the best fit in the log-log space (solid line), and adjust $Q/k$ 
such that $\tau_e=4.5$ Gyr for $a_{\rm b}/R=55$ (dashed lines). This gives $Q/k=2500$. To get some sense of the uncertainty
of this estimate, we repeat the same calculation for $a_{\rm b}/R=50$ and $a_{\rm b}/R=60$ obtaining 
$Q/k=4000$ and $Q/k=1500$, respectively. The density $\rho=1$ g cm$^{-3}$ was adopted here.}      
\label{tide}
\end{figure}

Figure \ref{tide} shows the circularization timescale for known TNO binaries. We find $Q/k \simeq 2,500$ 
such that $\tau_e \simeq 4.5$ Gyr for binaries with $a_{\rm b}/R=55$. Using $Q/k \simeq 2,500$ in Eq. (\ref{omega}) 
we can estimate how long it takes to reach a synchronous state starting from the critical rotation (Fig. \ref{synch}). 
As we will see below, the binary lifetime (i.e., the time interval before a binary becomes dissociated by 
collisions or encounters) depends on the outer disk mass $m_{\rm disk}$ and lifetime $t_{\rm disk}$. Adopting 
$m_{\rm disk}=20$ Earth masses and $t_{\rm disk}=10$-100~Myr from Nesvorn\'y et al. (2018), we find that binaries 
with $a_{\rm b}/R<30$-45 should become synchronous. This corresponds to the rotation periods up to 
400-750 hours. We thus see that tides are capable of producing very long rotational periods. The surviving 
binaries on Trojan orbits would evolve by tides over 4.5 Gyr and can become synchronous for $a_{\rm b}/R<80$ 
(Fig. \ref{omega}). To have $P_s>100$ h in a double synchronous state, $a_{\rm b}/R>12$. In summary, the spin 
periods $P_{\rm s}>100$ h observed among Jupiter Trojans could be obtained from dissociated binaries with 
$12<a_{\rm b}/R<30$-45 and/or surviving binaries with $12<a_{\rm b}/R<80$.

\section{Binary dissociation and the yield of slow rotators}

The outer disk lifetime is tied to Neptune's migration. If Neptune's migration started early (Ribeiro de Sousa
et al. 2020), the disk was short-lived (e.g., $t_{\rm disk}\sim10$ Myr) and there was less opportunity for 
collisions to dissociate binaries. If the disk was long-lived (e.g., $t_{\rm disk}\sim100$ 
Myr), more binaries would become unbound by impacts. Binaries were also dissociated by dynamical perturbations 
during planetary encounters that occurred as bodies evolved from $\sim$20-30 au to 5.2 au (Nesvorn\'y et al. 
2018, Nesvorn\'y \& Vokrouhlick\'y 2019). We determined the fraction of dissociated binaries using the methods 
described in Nesvorn\'y et al. (2018) and Nesvorn\'y \& Vokrouhlick\'y (2019). Here we first consider the case 
with $t_{\rm disk}=10$ Myr. Adopting $R=15$ km as a reference value, we find that $\simeq$60-80\% 
of binaries with $12<a_{\rm b}/R<30$ would be dissociated in this case (Fig. \ref{short}). As planetary encounters 
provide the dominant dissociation channel, the components of dissociated binaries should maintain their 
original, presumably slow spin. 

If we assume, for example, that $\sim$50\% of TNOs formed as binaries (meaning that $\sim 2/3$ of all individual 
TNOs were members of binaries; Noll et al. 2008, Fraser et al. 2017), $\sim 1/3$ of binaries had 
$12<a_{\rm b}/R<30$ and became double synchronous, and $\simeq 60$-80\% of binaries with $12<a_{\rm b}/R<30$ 
were dissociated (Fig. \ref{short}), then the expected fraction of slow rotators with $P>100$ h from 
dissociated binaries would be $\sim 2/3 \times 1/3 \times 0.7 = 0.16$. Here we conservatively assumed that 
all binaries with $a_{\rm b}/R>30$ became dissociated and contribute to singles. The fraction of slow rotators 
from dissociated synchronous binaries could thus explain observations (Szab\'o et al. 2017, Ryan et al. 2017).
In addition, $\sim$30\% of binaries with $12<a_{\rm b}/R<30$, and $<$20\% of binaries with $30<a_{\rm b}/R<90$,  
would have survived for $t_{\rm disk}\sim10$ Myr, suggesting that at least some slow rotators may be surviving  
binaries. This possibility can be tested observationally.

In long-lived disks, binaries are predominantly dissociated by impacts (Nesvorn\'y \& Vokrouhlick\'y 2019;
their Fig. 8). For example, $\sim$80-90\% of equal-size binaries with $R=15$~km and $12<a_{\rm b}/R<30$
would be dissociated for $t_{\rm disk}=30$ Myr. On one hand, the longer timescale would allow wider binaries 
to become synchronous and the larger dissociated fraction would increase the yield. On the other hand, 
a large number of impacts in long-lived disks would affect rotation (see below) and potentially remove any 
excess of slow rotators. This would have adverse implications for the overall yield of slow rotators. We thus 
see that the existence of slow rotators among Jupiter Trojans could be used to favor short-lived disks.

\section{Spin-changing collisions}

Rotation of a small body can change as a result of impact. We tested this effect using the Boulder collisional 
code (Morbidelli et al. 2009). We assumed that the slow rotators have $P_{\rm s}=100$-500 h initially and 
adopted the outer disk parameters from Nesvorn\'y \& Vokrouhlick\'y (2019). A large uncertainty in modeling the 
effect of impacts consists in coupling of the impactor's linear momentum to the angular momentum change 
of the target (e.g., Dobrovolskis \& Burns 1984, Farinella et al. 1992). This is because large and oblique
impacts, which are the most important for spin changes, can eject substantial amounts of material in the 
direction of projectile's motion. Thus, only a fraction $f<1$ of the projectile's momentum ends up 
contributing to target's spin. For $t_{\rm disk}=10$ Myr, we find that $\sim$55\% ($\sim$40\%) of initially 
slow rotators with $P_{\rm s}>100$ h and $D=30$ km end up with $P_{\rm s}<100$ h for $f=1$ ($f=0.3$). 

If the spin-changing impact happens {\it before} a binary is dissociated, the spin can be re-synchronized by 
tides. If the spin-changing impact happens {\it after} the binary dissociation event, however, there is no 
way back to slow rotation via tides. We thus see that spin-changing collisions should reduce the net yield 
of the mechanism described here. The reduction factor will depend on the poorly understood coupling 
parameter $f$ and $t_{\rm disk}$, with larger reductions occurring for stronger coupling and longer disk lifetimes. 
Future experimental work and impact simulations could help to quantify $f$ and help to disentangle effects 
of these two critical parameters. The spin-changing collisions become relatively rare {\it after} the 
implantation of bodies onto Jupiter Trojan orbits. Surviving binaries should therefore have ample time to 
tidally synchronize (and they do, as epitomized by the Patroclus-Menoetius binary).

\section{Caveats and outlook}

The proposed model hinges on a number of assumptions. We adopted the capture model of Jupiter Trojans 
from Nesvorn\'y et al. (2013), which is consistent with the orbital distribution and number of Jupiter 
Trojans, but other capture models are possible as well (e.g., Pirani et al. 2019). We assumed that 
the 100-km class TNO binaries with low binary eccentricities were circularized by tides and can therefore 
be used to infer the $Q/k$ parameter. Moreover, we assumed that the estimated value, $Q/k=1,500$-4,000, 
can be applied to small $D=15$-50 km binaries as well.

Goldreich \& Sari (2009) suggested that $k \sim 10^{-5} (R/{\rm km})$ for objects with rubble pile interior.
Adopting this scaling here, the $Q/k$ value inferred from circularization of tight TNO binaries would suggest 
$Q \sim 1$ for sizes considered here. This could mean that small TNOs can very efficiently dissipate the energy 
stored in the tidal distortion. For comparison, Goldreich \& Soter (1966) found $Q \sim 10$-500 for the 
terrestrial planets and moons of the outer planets. Our interpretation implies that the TNO binaries with 
$a_{\rm b}/R<55$ should have synchronous rotation. This can be tested by photometric observations. 

Additional uncertainty is related to the number of $12<a_{\rm b}/R<30$ binaries that formed in the massive 
trans-Neptunian disk at 20-30 au. We know from observations of cold classicals that the overall binary fraction may 
have been high (e.g., Noll et al. 2008, Fraser et al. 2017, Nesvorn\'y \& Vokrouhlick\'y 2019), but we do 
not know how high it was. These relatively tight binaries are difficult to detect observationally. 
To move forward, it would be desirable to identify more slow rotators among Jupiter Trojans and in other 
small body populations, and establish whether at least some of them are synchronous binaries.  

The model discussed here could also explain slow rotators in the asteroid and Kuiper belts. The fraction 
of very slow rotators with $P_{\rm s}>100$ h in the asteroid belt, and among Hildas in the 3:2 resonance 
with Jupiter, could be almost as high as the one found for Jupiter Trojans (e.g., Szab\'o et al. 2020, P\'al 
et al. 2020). The equal-size binaries (e.g., the Antiope binary; Merline et al. 2000) may have formed in situ 
in the asteroid belt or been implanted (Levison et al. 2009, Vokrouhlick\'y et al. 2016). The asteroid belt,
however, is thought to have experienced rather intense collisional evolution (Bottke et al. 2005), implying
very low chances of binary survival. Collisions have a profound effect on the spin distribution 
of asteroids as well (e.g., Farinella et al 1992). In this light, it is intriguing that (253) Mathilde
($D=53$ km) survived several large-scale impacts to end up with a very slow rotation ($P_{\rm s}=418$ h; 
Mottola et al. 1995).
 
\acknowledgements
The work of D.N. was supported by the NASA Emerging Worlds program. The work of D.V. was supported 
by the Czech Science Foundation (grant 18-06083S). W.M.G. acknowledges support from NASA through grant number
HST-GO-15143 from the Space Telescope Science Institute, which is operated by AURA, Inc., under NASA 
contract NAS 5-26555.

\clearpage

\clearpage

\clearpage

\clearpage
\begin{figure}
\epsscale{0.8}
\plotone{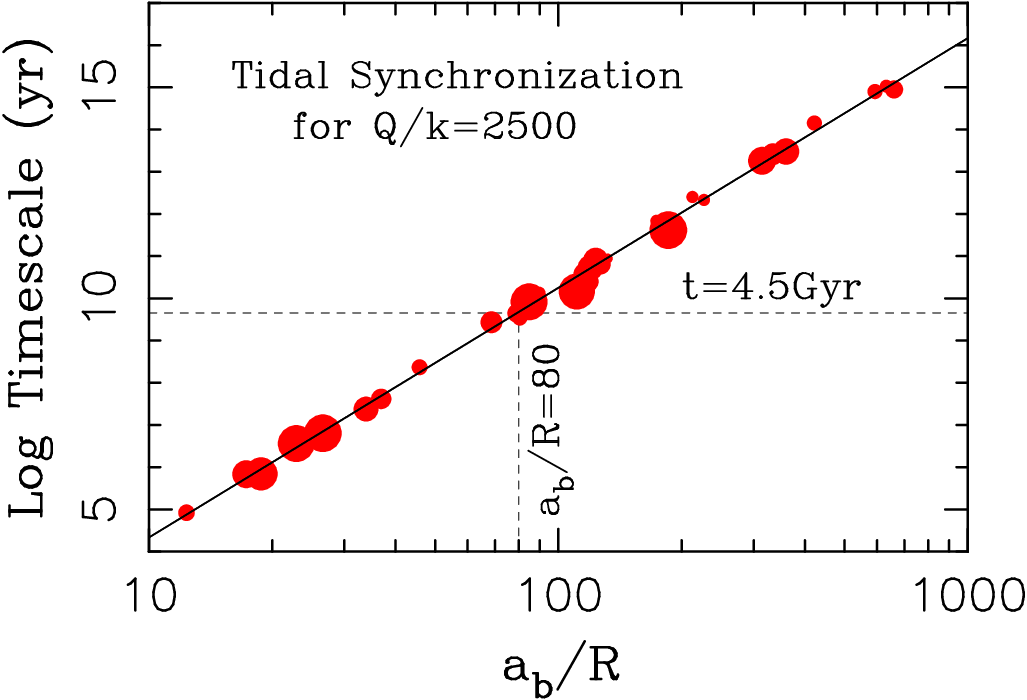}
\caption{The tidal synchronization timescale $\tau_\omega$ for $Q/k=2500$. We compute $\tau_\omega$ from 
Eq.~(\ref{omega}) for TNO binaries shown in Fig. \ref{circ}. The secondary spins are expected to be 
synchronized for up to $a_{\rm b}/R \simeq 80$ over the age of the solar system (dashed lines). Also, the 
synchronization timescale is $\tau_e=10$ Myr for $a_{\rm b}/R \simeq 30$ and $\tau_e=100$ Myr for 
$a_{\rm b}/R \simeq 45$. For the nearly equal-size TNO binaries the primary components become synchronized on only slightly 
longer timescales. The density $\rho=1$ g cm$^{-3}$ was adopted here.}      
\label{synch}
\end{figure}

\clearpage
\begin{figure}
\epsscale{0.8}
\plotone{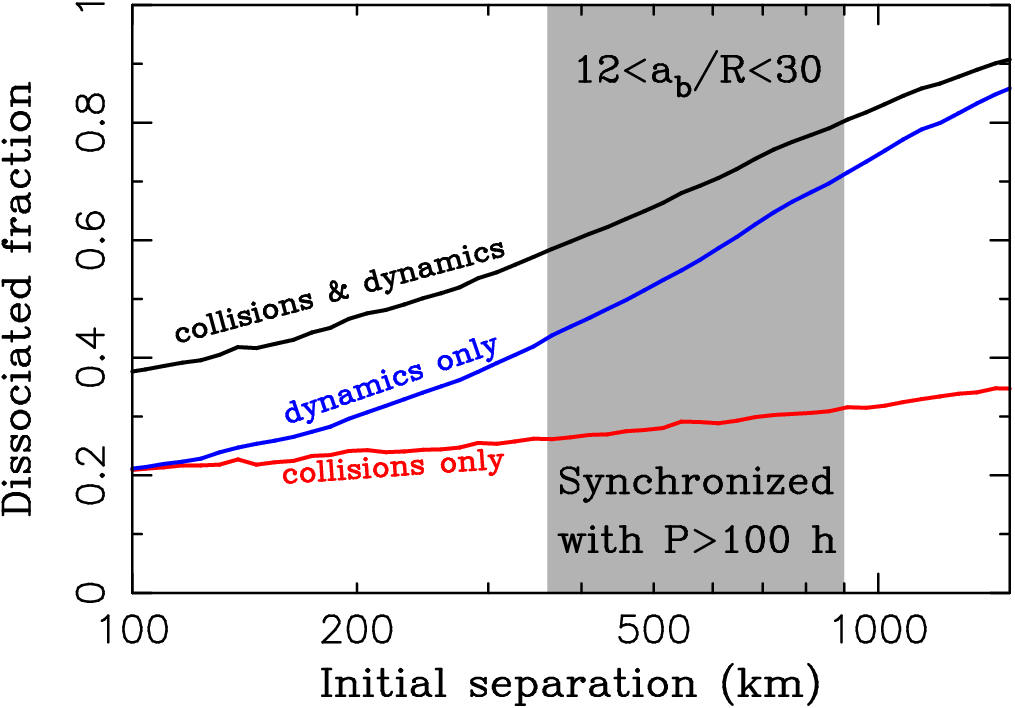}
\caption{The dissociated fraction of equal size binaries with $D=30$ km in the case of short-lived 
outer disk ($t_{\rm disk}=10$ Myr). Different lines show the dissociated fraction as function
of binary separation. The black line shows total fraction of dissociated binaries when both
collisions and planetary encounters are taken into account. The shaded area denotes binary separations 
where binaries are expected to become synchronized over 10 Myr and have $P_{\rm s}>100$ h.
In total, 60-80\% of these binaries become dissociated and 20-40\% survive for $t_{\rm disk}=10$ Myr.}
\label{short}
\end{figure}

\end{document}